\documentclass[11pt,a4paper]{emulateapj}
\newcommand\ms{M$_{\odot}$}
\slugcomment{Accepted to The Astrophysical Journal}

\shorttitle{Intergalactic Stellar Population from Mergers of E's}
\shortauthors{Gonz\'alez-Garc\'{i}a et~al.}
  
\begin{document}
  
\title{The Intergalactic Stellar Population from Mergers of Elliptical Galaxies with dark matter halos}
\author{A. Cesar Gonz\'alez-Garc\'{\i}a}
\affil{Departamento de F\'{\i}sica Te\'orica, Universidad Aut\'onoma de Madrid, 28049 Madrid, Spain}
\email{c.gonzalezgarcia@uam.es}
\author{Letizia Stanghellini}
\affil{National Optical Astronomy Observatory, 950 N. Cherry Av.,
Tucson, AZ  85719}
%\email{lstanghellini@noao.edu}
\and
\author{Arturo Manchado}
\affil{
Instituto de Astrof\'isica de Canarias, v\'{\i}a L\'actea s/n, 
La Laguna, E-38200 Tenerife, Spain}
%\email{amt@iac.es}

\begin{abstract}
We present simulations of dry-merger encounters between pairs of elliptical galaxies with dark matter halos. The aim of these simulations is to study the intergalactic stellar populations produced in both parabolic and hyperbolic encounters. We model progenitor galaxies with total-to-luminous mass ratios  ${\rm M_T/M_L=}$ 3 and 11. The initial mass of the colliding galaxies are chosen so that ${\rm M_1/M_2=1}$ and $10$. The model galaxies are populated by particles representing stars, as in Stanghellini et al. (2006), and dark matter. 
Merger remnants resulting from these encounters display a population of unbounded particles, both dark and luminous. The number of particles becoming unbounded depends on orbital configuration, with hyperbolic encounters producing a larger luminous intracluster population than parabolic encounters. Furthermore, in simulations with identical orbital parameters, a lower ${\rm M_T/M_L}$  of the colliding galaxies produces a larger fraction of unbounded luminous particles. 
For each modeled collision, the fraction of unbounded to initial stellar mass is the same in all mass-bins considered, similarly to what we found previously by modeling encounters of galaxies without dark halos. The fraction of intergalactic to total luminosity resulting from our simulations is $\sim 4\%$ and $\sim 6\%$ for dark-to-bright mass rations of 10 and 2 respectively. These unbounded-to-total luminous fractions are down from $17 \%$ that we had previously found in the case of no dark halos. 
Our results are in broad agreement with intergalactic light observed in groups of galaxies, while the results of our previous models without dark halos better encompass observed 
intracluster populations. We suggest a possible formation scenario of intergalactic stars.

\end{abstract}


\section{Introduction}

The population of gravitationally unbounded intergalactic (IG) stars has been recently observed in different environments, such as in poor groups (Feldmeier et~al.~2004, Castro-Rodr{\'{\i}}guez et~al. 2003), compact  tidal groups (White et~al. 2003), and nearby galaxy clusters (Ferguson et~al. 1998; Durrell et~al. 2002; Feldmeier et~al. 2004, Gerhard et~al. 2005), as well as in higher redshift clusters (Zibetti et~al. 2005, Zibetti 2008). 
The amount of observed IG starlight seems to correlate with the environment, with more IG starlight found in populated clusters, and toward the cluster centers.
Zibetti (2008) studied 683 clusters of galaxies,  and found that the fraction of diffuse to total starlight decreases with the distances of the
unbounded stars from the galactic centers, a result that is also supported by Murante et al. (2004)'s predictions.

Several mechanisms have been proposed to explain the origin of the IG starlight. Muccione \& Ciotti (2004) associated the presence of IG stars to the evaporation of galaxies due to tidal interactions with other cluster members, and with the cluster potential well. Conroy, Wechsler \& Kravtsov (2007) showed that more than $80\%$ of the stars from tidally disrupted sub-halos contributes to the IG starlight. 

As an alternative to tidally produced intracluster stars, we (Stanghellini et al. 2006, hereafter Paper I) have shown that
dry mergers could produce an IG stellar population compatible with what has been observed to date in the intracluster. Our models of Paper I show that  during the merging stage a number of stars from the progenitor systems become unbounded, and the IG to initial starlight obtained was broadly consistent with the observed intracluster star counts. 

Murante et al.~(2007) analyzed a large number of clusters from a cosmological simulation. They find that most of the stars in the IG stellar population come from the formation of the brightest cluster galaxy, and that most stars of this diffuse component come from merging events and only a minor fraction of the stars derive from tidal interactions.

 In this paper we extend our models of interacting galactic pairs, whose components have dark matter halos. There is now ample evidence that disk galaxies should host a fair amount of dark matter to account for their kinematics (see, van Albada et al. 1985 and references thereafter). A further indication that elliptical galaxies have dark halos comes from cosmological simulations, and the hierarchical merging paradigm (see, e.g. O\~norbe et al. 2007). It is thus appropriate to repeat the simulations of colliding pairs of elliptical galaxies as in Paper I, this time assuming that the initial galaxy models have dark halos. We perform our simulations
keeping all other assumptions unvaried with respect to those of Paper I, so we can directly compare the results
from both sets of simulations.

In Section 2 we introduce our models and all the assumptions made for the galaxies, the dark halos, the stellar populations, and the collisions. Section 3 illustrates the results from the dry-merger simulations, with particular attention to the production of the IG stellar population. The discussion, including comparing the new, dark-haloed models with those of
Paper I, is in $\S$4. Section 5 presents the conclusions of our study.

\section{Models}

We have performed a series of simulations of dry mergers of pairs of spheroidal galaxies with dark matter halos. We chose the mass ratio between the 
two colliding galaxies to be  ${\rm M_2/M_1}$=1 and 10 \footnote{Hereafter the suffix $_1$ refers to the less massive galaxy of the merging pair, $_{\rm L}$ refers to the luminous mass, $_{\rm H}$ refers to the dark halo mass and $_{\rm T}$ to the total mass of an individual model, ${\rm M_T=M_H+M_L}$. Unless otherwise noted, M is the total mass.}.

\subsection{Galaxies, Initial conditions}

The initial conditions of the galaxies forming the colliding pairs are similar to those described by Gonz\'alez-Garc\'{\i}a \& van Albada 
(2005), who performed a study of the encounters between two spherical 
systems with dark matter halos. 
We use the isotropic spherical Jaffe's (1983) approach to model the initial conditions of the luminous matter, and the Hernquist (1990) model for the dark matter halos.

The potential for a Jaffe model is:

\begin{equation}
	\phi_{\rm L}(r) = \frac{GM_{\rm L}}{r_{\rm J}} \ln \left(\frac{r}{r+r_{\rm J}}\right),
\end{equation}
where G is Newton's constant, and $r_{\rm J}$ is the half-mass radius of the luminous component.

The potential of the Hernquist model is:

\begin{equation}
	\phi_{\rm H}(r) = -\frac{GM_{\rm H}}{r+a},
\end{equation}
where $a$ is the scale length used.
The half-mass radius of the dark halo component is equal to $(1 + \sqrt{2}) a$.
We combined the Jaffe and the Hernquist model to reproduce the haloed galaxy. We obtained a two-parameter family of galaxies that depends on the luminous-to-dark mass ratio, ${\rm M_L/M_H}$, and on the half-mass radius ratio,

\begin{equation}
	r_{\rm L_{1/2}}/r_{\rm H_{1/2}} =\frac{r_{\rm J}}{(1+\sqrt{2})a}.
\end{equation}

The distribution function of the combined system is :

\begin{equation}
	f_{\rm T}(E) =f_{\rm L}(E)+f_{\rm H}(E).
\end{equation}

An algorithm yielding $f_{\rm L}(E)$ and $f_{\rm H}(E)$ along the two-parameter family was developed by P.A.H. Smulders \& M. Balcells (1995, unpublished, 
see Gonz\'alez-Garc\'{\i}a \& van Albada 2005 for a detailed description of the algorithm). 

In the following, we adopt non-dimensional units with $G=1$ for Newton's constant of gravity. The theoretical 
half-mass radius of the Jaffe model, $r_{\rm J}$, and the total luminous mass of the fiducial galaxy 
in each run are also set to 1. The models may be compared with real galaxies using the 
following scaling, similar to the one employed in Paper I:

\begin{equation}
	      {\rm  [M]} = 4\times10^{11} \; {\rm M_{\odot}}, \\
\end{equation}
\begin{equation}
		{\rm [L] }= r_{\rm J} = 5 \;{\rm Kpc}  ,\\
\end{equation}
\begin{equation}
		{\rm [T]} =  8.34\times10^6 \;{\rm yr}. \\
\end{equation}

By adopting these units, the velocity unit is:

\begin{equation}
		[v] =  587\; {\rm km \; s^{-1}}. \\
\end{equation}

The initial systems are spherical and non-rotating, with an isotropic velocity distribution. In this regard they are similar to the set of spherical isotropic systems 
studied in Paper I.

The observed amount of dark matter in elliptical galaxies is controversial. Elliptical galaxies lack the reliable kinematic tracers that allow the detailed dynamical studies of spiral galaxies. Recently there have been improvements in recovering the presence of dark matter in elliptical galaxies, such as in the X-ray study by Humphrey et al. (2006), via gravitational lensing (Treu \& Koopmans 2004, Mandelbaum et al. 2006), and through stellar dynamics (Thomas et al. 2007).  Chakrabarty \& Raychaudhury (2008) have applied globular cluster kinematics to estimate a ratio of $\sim$10 for the dark to luminous mass in the elliptical galaxy NGC 4636. Romanovsky et al. (2003), from observations of the kinematics of planetary nebulae in the outer parts of elliptical galaxies set a firm upper limit to the amount of their dark matter, although it has been noted that their results may be model dependent (Dekel et al. 2005). 

In the present models, we have initially used a range of $M_{\rm H}/M_{\rm L}$ values, and calculated the circular velocity curve for those realizations (bearing in mind that we use a composite of
the Jaffe and Hernquist models), aiming at a flat-topped curve for a significant radial range. We obtained the best match for $M_{\rm H}/M_{\rm L} = 2$ and $10$, and a halo scale length $a=2$ %(for $r_{\rm J}=1$), 
which yield component ratios of $M_{\rm T}/M_{\rm L}$=3 and 11. These ratios are consistent with the observations of Mandelbaum et al. (2006). The models presented here can be viewed as extrapolations of the Paper I models, where no dark matter halo was included (i.e., $M_{\rm T}/M_{\rm L}$=1). Dark halos have been populated with $5 \times 10^5$ particles for all galaxy models, except the smallest galaxy in run $10hP$ which has ten times less particles.

The projected surface density of luminous material of such models presents a slope that decreases roughly as $R^{1/4}$, 
which makes it a suitable representation for elliptical galaxies, although the central parts 
present a cusp. 

In Table~\ref{tab:1} we summarize the characteristics of the initial models. Column (1) gives the run
identification code, where the initial number indicates the mass ratio of the interacting galaxy pair, the small letter denotes 
the impact parameter ({\it h} is for head-on impact), and the capital letter denotes 
the energy of the orbit, with {\it P} for parabolic and {\it Z} for zero energy at infinity (or hyperbolic) orbit;
column (2) gives the total-to-luminous mass ratio of the initial galaxies; column (3) gives the mass ratio of the colliding galaxies (where the masses represent the total masses for each galaxy);  column (4) gives the initial separation $r_i$, in model units;
column (5) gives the initial relative velocity of the galaxy pair, $v_i$, also in model units;
and columns (6) and (7) give respectively the impact
parameter, and the orbital energy of the initial setup. 

\begin{table}
\begin{minipage}{85mm}
\caption{Galaxy input parameters \label{tab:1}}
\begin{tabular}{lrrrrrr}
\hline
{\bf Run} & {\bf M$_{\rm T}$/M$_{\rm L}$} & {\bf M$_2$/M$_1$}&{\bf $r_i$}&{\bf $v_i$}&{\bf b} &{\bf E$_{\rm orb}$} \\
\hline
$1hP$ & 11& 1 & 40 & 0.316& 0 & 0 \\
$1hZ$ & 11& 1 & 40 & 1.048& 0 & 0.250 \\
$10hP$ & 11& 10 & 19.48 & 0.336&0 & 0\\
$1hP3$ & 3& 1 & 40 & 0.316& 0 & 0 \\
\hline
\end{tabular}
\end{minipage}
\end{table}
%\vspace{0.75cm}

To appropriately scale the models with mass other than 1, we impose that the luminous mass of initial systems must lie on the fundamental plane of galaxies which, according to J\o rgensen et al. (1996), gives:

\begin{equation}
\log{\rm L} \simeq 0.78 \log {\rm M_T} + constant.
\end{equation}
and we use Fish's (1964) law to scale the radii of the luminous and dark components of the two systems.

Model $1hP$ is an equal mass encounter (${\rm M_1=M_2}$) with the interacting galaxies set on a parabolic orbit. 
The centers of the two galaxies are placed at an initial distance of $4~ R_1$,
where $R_1=10~ r_{\rm J}$ is the galaxy radius. 
Model $1hZ$  is an equal mass encounter with the progenitors on a mildly hyperbolic orbit, with all the other parameters as in model $1hP$.
Model $10hP$ is an encounter between two galaxies with a mass ratio of 10 on a radial 
parabolic encounter, initially placed at a distance ${\rm 3~ R_1 +R_2}$, where ${\rm R_1}$ and 
${\rm R_2}$ are respectively the radii of the less and the more massive galaxies.
Following the definitions above, ${\rm R_2= 10~ r_J}$ and ${\rm R_1=3.16 ~r_J}$. 
Finally,  Model $1hP3$ has the same orbital parameters as model $1hP$, but with different total to luminous components ratio of the progenitor system, ${\rm M_T/M_L}$=3.

%The choices of initial conditions for the galaxies are
%astrophysically adequate to represent the observations. 
The relative range of the mass of the galaxy pairs chosen for the 
simulations,
${\rm M_2/M_1}$=1 or 10, are the extremes of the observed merging galaxy 
sample by van Dokkum (2005).

\subsection{Stellar components}

The luminous components of our galaxies models have been populated by particle stars that follow Salpeter's (1955) initial mass function (IMF), $\Psi(m)\propto m^{-2.35}$, and are distributed into three 
representative mass bins. A detailed discussion of this implementation is given in Paper I. The population of each mass bin corresponds to the 
integration of the Salpeter's mass function in that bin, scaled
to the entire population considered, and ignoring stars 
outside that mass range. The luminous particles in our fiducial galaxy model have mass distributions such as that, after accounting for the Salpeter's IMF, 
the total galaxy luminous mass is equal to unity.

In Table~\ref{tab:2} we summarize the luminous population of the input fiducial galaxy. 
Column (1) gives the mass bin in solar masses, column (2) gives the mass fraction in that mass bin, 
column (3) gives the number of luminous particles in the bin, 
and column (4) the (dimensionless) mass of particles in each bin. 

In runs $1hP$, $1hZ$ and $1hP3$ both galaxies have number of particles per mass bin as in Table~\ref{tab:2}. 
The stellar content of Table~\ref{tab:2} also represent the more massive galaxy in run $10hP$,
while the less massive galaxy ($M_{\rm L}=0.1$) has an initial set up with ten 
times less particles in each mass bin, but with the same mass per particle.

\begin{center}
\begin{table}%{lrrr}
\begin{minipage}{85mm}
\caption{Stellar input \label{tab:2}}
\begin{tabular}{lrrrr}
\hline
{\bf bin [\ms]} &{\bf $\Phi_{bin}$}&{\bf N$_{particles}$} &{\bf m$_{particle}$ } & {\bf  m$_{particle}$ [\ms]} \\
\hline
0.85 -- 1.4 & 0.5151& 156060 &  $3.3\times10^{-6}$ & $1.32\times10^{5}$\\
1.4 -- 3.0 & 0.3443& 52121&	$6.6\times10^{-6}$ & $2.64\times10^{5}$\\
3.0 -- 8.0 & 0.1406& 7050&	$2.0\times10^{-5}$ & $8.00\times10^{5}$\\
\hline
total& 1& 215231& 1\\ 
\hline
\end{tabular}
\end{minipage}
\end{table}
\end{center}
%\vspace{0.75cm}

\subsection{Integration method and stability}

We have used the parallel treecode {\small GADGET-2} (Springel 2005) on the MareNostrum Super Computer at the BSC. 
Softening was set to $\sim 1/10$ of the half-mass radius of the smallest galaxy ($\varepsilon = 0.075$). The tolerance parameter was set to $\theta=0.8$. Quadrupole terms were 
included in the force calculation and the maximum time step was set to $1/50$ of the half-mass crossing time.

We have checked the stability of our input initial model for $33$ half-light crossing times (see formula 2.40 in Binney \& Tremaine 2008, page 64; $\sim 0.5$ Gyrs). 
The test shows that the system relaxes for about 6 time units and remains stable thereafter. The inner parts present a shallower profile at the end of the relaxation time. 
This initial relaxation is due to the presence of the particle softening in the code. The stability runs are comparable to that performed for ten times less particles by Gonz\'alez-Garc\'{\i}a \& van Albada (2005) with small differences in the inner parts due to the smaller softening used in the present simulations.

The merged system was evolved for more than $10$ dynamical half-light crossing times of the merged body after 
merging, 
to allow the system to relax (reach virialization). Conservation of energy is good in 
all the runs; energy variations are less than $0.5$ per cent.

\section{Results}

The evolution of the runs and the final states are quite similar to those already described by Gonz\'alez-Garc\'{\i}a \& van Albada (2005), we refer to the interested reader to that paper for a detailed description of the merger stages. We present a brief description here.

Galaxies in simulation $1hP$ evolve towards a first passage through pericenter at $t=80$ (in model units of time). At this first encounter the systems exchange energy, the halos of the two galaxies already merge, and a fraction of both dark and luminous matter gains enough energy to escape. %The luminous objects remain on a bound elliptical orbit, leading to a final merger at approximately $t=150$. 
A similar situation is found in simulation $10hP$, although with different timescales. Galaxies in simulation $1hZ$, which has the highest initial orbital energy, merge after the first encounter and thus there is only a single release of particles.  Simulation $1hP3$ shows a different behavior due to the lower relative content of dark matter. The two systems meet at $t\sim80$, but the halos do not merge at this first encounter. A number of particles of the two systems, both dark and luminous, gain enough energy to escape. The two system meet again at $t\sim150$ and finally merge, again exporting some particles (dark and luminous) to the IG population. The simulations are stopped at time $t=200$, i.e. $t \sim 1.67 {\rm Gyrs}$.

\begin{figure*}

\plotone{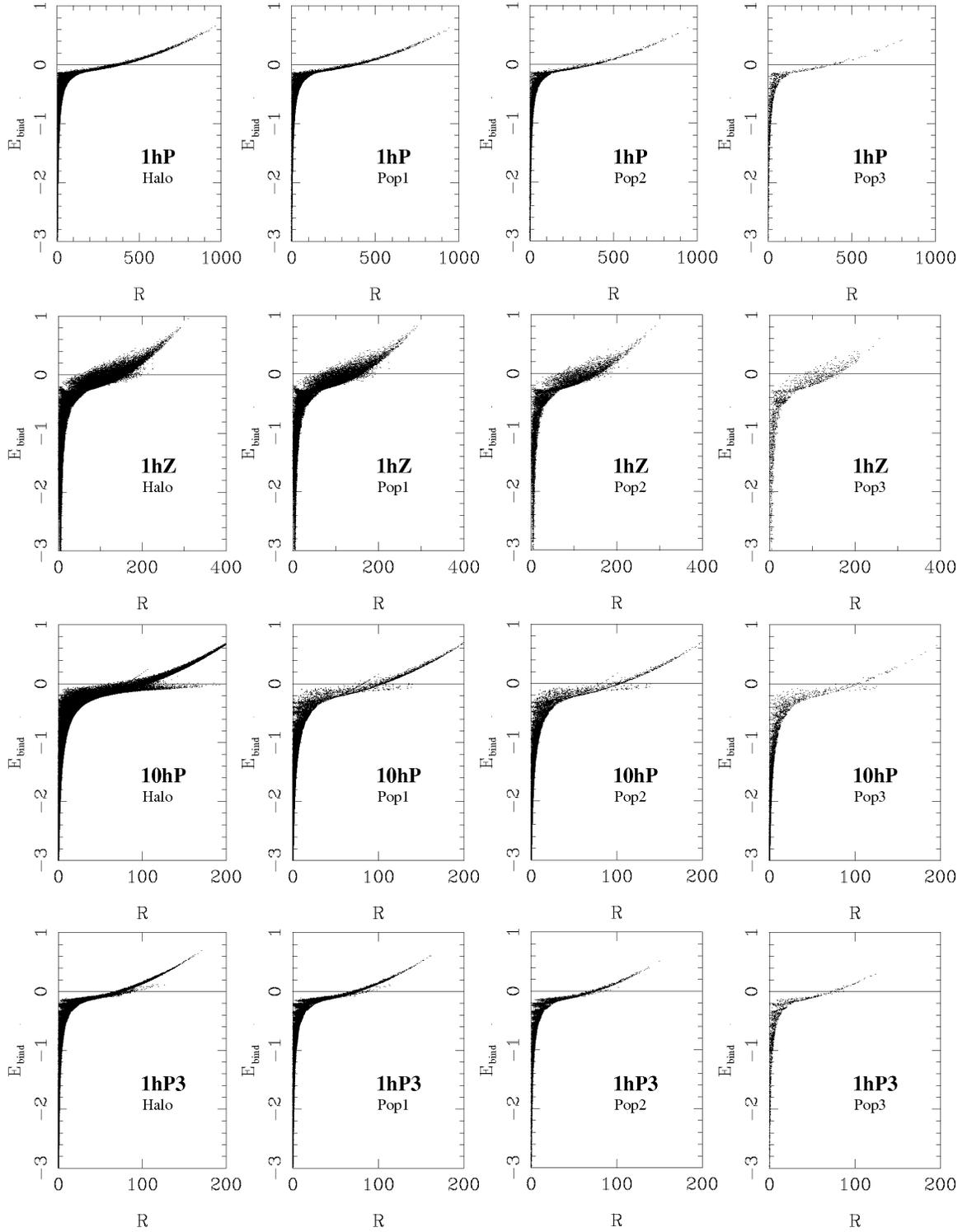}
%\plotone{fig_2.eps}
%\plotone{fig_3.eps}
%\plotone{fig_4.eps}
\caption{Binding energy vs. distance from the merger's center of models $1hP$ (top row), $1hZ$ (second row from the top), $10hP$ (third row from the top),  $1hP3$ (bottom row). Columns contains, respectively from left to right, dark matter particles, low-mass population of luminous particles, intermediate-mass population of luminous particles, and high-mass populations of luminous particles. Units are in the internal system of units, see text for details. \label{fig}}

\end{figure*}

Figure~\ref{fig} shows the final distribution of the various types of particles in our encounter simulations; in each panel the x-axis represents the distance from the merger center, R, and the y-axis the binding, ${\rm E_{bind}}$,  in dimensionless model units. We plot the results of different particles: from the left to the right panels: dark matter particles, then low-mass, intermediate-mass, and high-mass luminous particles are plotted.    The dark and luminous particles show similar qualitative behavior. When the galaxy systems pass through the pericenter, 
particles gain kinetic energy and may end up having a positive binding energy. 
Simulation $1hZ$ presents a larger number of particles with positive binding energy than all other types of encounters. Simulations $1hP$ and $1hP3$ show a moderate number of expelled particles. Finally, simulation $10hP$ shows very small number of particles with positive binding energy.

\begin{figure}

\plotone{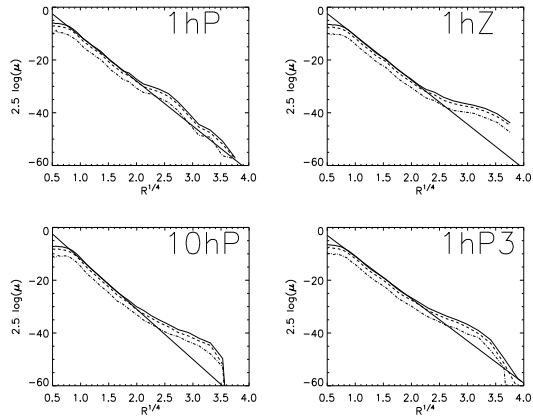}

\caption{Surface density vs. $R^{1/4}$ for the different simulations. From top to bottom, respectively, we show $1hP$,  $1hZ$,  $10hP$, and $1hP3$. The different lines are the results for the different stellar populations. Low-mass population: Solid line; intermediate-mass population: Dashed line;  and high-mass population: dotted-dashed line. A de Vaucoleurs profiles is shown as a thin solid straight line. Units are in the internal system of units. For details, see text. \label{fig1}}

\end{figure}

In Figure~\ref{fig1} we plot the logarithm surface density calculated in annuli of increasing radius from the merger's centers versus ${\rm R^{1/4}}$, where R is the distance from the merger center. The different lines represent the total surface density (solid) and the surface density for each of the luminous populations (see caption). 

The intermediate regions present a profile close to a de Vaucoleurs profile (see full straight line).  The innermost ($R^{1/4} < 0.7$) and outer regions deviate clearly from this profile. The inner parts are affected by the fact that our initial models have a core-like shape after the relaxation run. The outer regions deviate from the de Vaucouleurs profile due to the escaping IG material in a similar fashion as it was explained in Paper I. However, this deviation is less pronounced in the present models given the lower fraction of escaping particles. Deviations from the de Vaucouleurs' slope affected by the IG particles in the outer regions are important outside $R^{1/4}\sim 2-2.5$. These values are in agreement with those found in Paper I. The surface density excess with respect to a de Vaucouleurs fit (solid thin line) is larger for model $10hP$ than for model $1hP$ at large radii. This is due to the deposition of most luminous particles of the small galaxy at these large radii. In all models and at all radii the different populations contribute similarly to the IG population. 

In each of the proposed simulations, a fraction of particles escapes the main galaxian potential.  in Table 3 we give the resulting IG population produced in each run;
column (1) gives the run code; column (2) gives the mass bin; column (3) gives the fraction of unbounded luminous mass (with respect to total luminous mass); column 
(4) gives the fraction of the luminosity of unbounded particles, $L_u$, with respect to total luminosity, $L_T$ and (5) gives the fraction of the luminosity of unbounded particles versus the luminosity of the merger remnant, $L_m$. We also give, for each run and in column (3), the mass of the total (luminous and dark) unbounded mass as a fraction of the total initial mass.

By examining Table 3 we note that the fraction of gravitationally unbounded mass is higher in the total than in the luminous matter. This is occurring because
the dark halos extend further out than the luminous parts of the galaxies, 
thus the presence of the dark halo potential places the luminous matter deeper into the potential well. The fraction of total unbounded mass 
is roughly the same in runs $1hP$ and $10hP$, thus it is independent on the initial mass ratio. It is on the other hand higher
in $1hP3$ due to the different internal structures and relative energies while varying the total mass of the dark halo. The situation is different for $1hZ$, where the larger available energy results in 12$\%$ of the mass being expelled by the merging process.

The luminous mass expelled from the encounters of galaxies with dark halos in Table 3 is less than what was found in the models without  dark halos (i.~e., Paper I):  we find that 
the modeled stellar-to-total unbounded mass after the parabolic encounters of galaxies with dark halos is about 15 times smaller than in models without dark halos (it goes down to 10 times for the hyperbolic encounters).

Gonz\'alez-Garc\'{\i}a \& van Albada (2003) and Nipoti et al. (2003) demonstrated that a few merger events between systems on the FP do not destroy such relation (although see Nipoti, Treu \& Bolton 2009). The merger remnants in our sample must obey a law like ${\rm log~\rm L \simeq ~0.78 ~log ~M_T}$ + constant (J\o rgensen et al. 1996; equation 9). In the present study we have that $\rm M_{\rm T}=\rm M_{\rm L}+\rm M_{\rm H}$. Now, if we introduce this in Equation 9, we get: 

\begin{equation}
\log \frac{\rm L_1}{\rm L_2}=0.78\log \frac{\rm M_{\rm L1}}{\rm M_{\rm L2}} + 0.78 \log \frac{(\rm M_{\rm T1}/\rm M_{\rm L1})}{(\rm M_{\rm T2}/\rm M_{\rm L2})}
\end{equation}

We must translate the mass of our stellar particles into light. Our simulations are gravity-only simulations and we do not have their luminosity. To convert our stellar masses in luminosity we use the relation ${\rm M_T/M_L\propto M_T^{0.17}}$ from Hyde \& Bernardi (2009) combined with equation 9. Both the progenitor and the merged systems are on the FP relation, but the unbounded material is not. So, at a first step we want to compute the following: ${\rm L_u/L_m = (L_1+L_2-L_m)/L_m = (L_1/L_m) + (L_2/L_m) -1}$. Using the above mentioned formulas we get:

\begin{equation}
\rm \frac{L_u}{L_m} =   (\frac{M_1}{M_m})^{0.94} + (\frac{M_2}{M_m})^{0.94} - 1
\end{equation}

And we can derive the fractions, which are included in Table~\ref{tab:3}. Finally, we find that the ratio of luminous unbounded to total material $\rm L_u/L_T = L_u / (L_u+L_m) = 1 /[(L_u+L_m)/L_u)= 1 / (1 + (L_m/L_u))$, also included in Table~\ref{tab:3}. 

\begin{center}
\begin{table}%{lcrrrr}
\begin{minipage}{85mm}
\caption{Results \label{tab:3}}
\begin{tabular}{lcrrr}
\hline
{\bf Run}&{\bf bin [\ms]}&{\bf M$_{\rm u}$ ($\%$)}&{\bf L$_{\rm u}$/L$_{\rm T}$}&{\bf L$_{\rm u}$/L$_{\rm m}$}\\
&($M_{\odot}$)&($\%$)&($\%$)&($\%$)\\
(1)&(2)&(3)&(4)&(5)\\
\hline
%\startdata
$1hP$&	0.85 -- 1.4&	  0.33&      \\
&	1.4 -- 3.0&	  0.34&      \\
&	3.0 -- 8.0&	   0.31&     \\
&	total lum&	  0.33& 4.37 &4.57\\
&  Total & 2.06 & \\
\hline
$1hZ$&	0.85 -- 1.4&	  2.07&     \\
&	1.4 -- 3.0&	  2.04&     \\
&	3.0 -- 8.0&	  1.87&     \\
&	total lum&	  2.04& 5.92 &6.29\\
&  Total & 12.15 & \\
\hline
$10hP$&	0.85 -- 1.4&	  0.63&     \\
&	1.4 -- 3.0&	  0.60&     \\
&	3.0 -- 8.0&	   0.52&    \\
&	total lum&	  0.60& 2.38&2.43\\
&  Total & 2.41 & \\
\hline
$1hP3$&	0.85 -- 1.4&	  2.36&     \\
&	1.4 -- 3.0&	  2.22&     \\
&	3.0 -- 8.0&	  2.38&     \\
&	total lum&	  2.32& 6.17&6.58\\
&  Total & 3.25 & \\
\hline

%\enddata
\end{tabular}
\end{minipage}
\end{table}
\end{center}
%\vspace{0.5cm}

\section{Discussion}

In reviewing all our merging models we find that he total amount of the IG starlight  is determined mainly by two factors: the amount of dark matter in the colliding galaxies, and the energy of the encounter.
The first factor is clearly accounted for when  we look at simulations $1hP$ and $1hP3$ from this paper and model $1hP$ in Paper I :
The fractions of IG to total luminosity are $\sim 4 \%$, $\sim 6 \%$ and $17 \%$ for dark mass amounts of, respectively, 10, 2, and 0 times the amount of the luminous mass. 
On the other hand, increasing the mass ratio of the colliding galaxies (model $10hP$) does not produce a significant increased IG population: The unbounded luminosity produced in this model  is small when compared with the other simulations. This is reflecting the fact that most of the IG material originates, in this case,  from the low-mass galaxy. This also implies that satellite-like or minor mergers should be at least between two and ten times more frequent to get a similar  IG stellar population than that produced by major mergers. 

We can compare the present results and those in Paper I with other models and simulations. The semi-analytic predictions of Purcell, Bullock \& Zentner (2007) claim that the intrahalo light should rise from $0.5$ to $20\%$ from galaxy halos to poor groups, while it should be between $20-30 \%$ for rich groups and clusters. Such numbers for the IG light in poor groups are in rough agreement with our results for the simulations presented here for models with a dark halo, while their numbers for clusters are in closer agreement with our results from Paper I.

Observations might indicate an increase of the relative IG to total starlight as the mass of the group increases. Loose groups show fractions of the order of less than 2$\%$ (Castro-Rodr\'{\i}guez et al.~2003) while, clusters like Virgo present fractions around 5-10$\%$ (Arnaboldi et al.~2003) and massive clusters  fractions larger than 10$\%$ (Feldmeier et al.~2004), although Zibetti et al.~(2005) find that the fraction of  IG to total starlight is almost independent on the cluster richness.

Aguerri \& Gonz\'alez-Garc\'{\i}a (2009) show that a galaxy entering a dense environment, such as a cluster, it is stripped away from the outer parts (mostly dark matter), due to tidal stripping with the potential well of the cluster as well as due to galaxy harassment. The harassed galaxy has, as a consequence, a lower dark-to-luminous mass ratio. This transformation is more severe in dense environments where faster encounters are likely. Thus, denser environments should host systems with lower dark-to-luminous ratios. Conversely, groups should host systems with larger dark-to-luminous ratios. 

In summary, in loose groups the potential well of the group is not large enough to strip the galaxy from all its dark matter halo. This scenario is likely to present mergers of galaxies with a fair amount of dark matter involved, like the ones investigated in this paper. The shielding effect of the dark halos should then prevent a large fraction of IG stars in these environments. Dense groups and rich clusters, on the contrary, are environments where tidal stripping and tidal harassment with the halo potential should affect primarily the outer parts of the galaxies, stripping away the halo; this might have the effect of lowering the initial relative velocities and the components dark-to-luminous ratio, and the galaxy pairs might undergo encounters such as those presented of $1hP3$ in this paper, or those presented by Paper I. This could explain why these environments present a larger IG light fraction.

The parameter space explored here is naturally limited. In this sense, this paper and Paper I must be viewed as a first attempt. We test the assumption that most of the IG population come from dry-mergers, so gaseous mergers involving disk galaxies, star formation, and re-processing are excluded from our modeling. Mergers between gas-free disk (S0-like) galaxies have recently been explored  (Wu \& Jiang 2009) with results compatible with those presented here. 

Other caveats of our approach could be that we do not model the collisions arising from cosmological simulations, and thus we do not have fully cosmological conditions. Our dark matter halo is a Hernquist profile, which is a cored halo profile. However, cosmological simulations tend to produce cuspier density profiles (see e.g., Navarro, Frenk \& White 1996). We have run a test simulation with an NFW dark matter halo profile, constructed with the software described by Widrow \& Dubinski (2005), with initial conditions identical to run $1hP$. The fractions of unbound light are: $\rm L_u/L_m = 4.43 \%$ and $\rm L_u/L_T = 4.24 \%$. These figures are roughly similar to those found in runs with shallower inner profiles. Altough the NFW is deeper at the central parts, we must note that the unbound matterial is extracted from the outer layers of both colliding systems where the potential well is shallower.

Also, in clusters the matter distribution in satellite halos is modified and would be better tested with other profiles like those proposed by Springel et al. (2008). Such problems should be explored in the future using a re-simulation of a cluster or group area from a fully cosmological simulation. Finally, we must note that the IG light could have been pre-processed in groups and later accreted to a cluster, contributing with their share of already unbound stars (see, e.g. Rudick, Mihos \& McBride, 2006).

\section{Conclusions}

We have modeled dry-mergers  between pairs of elliptical galaxies populated with luminous particles, surrounded by dark matter halos and predicted the extent of the intergalactic stellar populations produced in these encounters. These simulations modeled different orbital parameters, mass ratios, and varying initial dark-to-luminous mass ratios. These simulations are an extension of those presented in Paper I, which were without dark matter halo.

The merger process injects orbital energy into the final system. Some particles, both luminous and dark,  gain enough energy to become unbounded, populating the modeled intergalactic space. The resulting fraction of unbounded to initial dark matter is larger than the corresponding fraction of luminous particles. 
We also find that the presence of a dark matter halo results in a lower amount of free-floating luminous particles, translating into an expectation of more intracluster
or intragroup stars in low-dark matter galaxy encounters.

An interesting point would be to know when particles become unbound and which dynamical process (violent relaxation, tidal stripping at pericenter...) is the dominant one. However, this is out of the scope of the present article and will be investigated in forthcoming papers.
Finally, 
the fraction of IG starlight observed in the different simulations presented in this paper are in rough agreement with those reported in analytical 
models, and in observations of poor galaxy groups. The fractions presented in Paper I are in closer agreement with those observed in the intracluster.

We conclude that dry merging is a viable process to explain the IG unbound stellar population in many specific cases of low group density.

\acknowledgements

This work was partially supported by the MCyT (Spain) grant AYA2006-15492-C03-01 from the PNAyA, the regional government of Madrid through the ASTROCAM Astrophysics network (S­0505/ESP­0237), and by NOAO. We thank the BSC in Barcelona for computing facilities at the Mare Nostrum supercomputer and at the La Palma node, and 
the Centro de Computaci\'on Cient\'{\i}fica (UAM, Spain). We would like to thank the anonymous referee for interesting comments which helped to improve the quality of the paper.

%
%\newpage

\end{document}